\documentclass[10pt]{article}
\usepackage{cite}
\usepackage{epsfig}
\usepackage[dvips,usenames]{color}
\usepackage{color}
\usepackage{rotating}
\usepackage{amsmath,amssymb,amsfonts}
\usepackage{latexsym}
\begin{document}
\setcounter{section}{0}
\setcounter{equation}{0}
\setcounter{figure}{0}
\setcounter{table}{0}
\setcounter{footnote}{0}
\begin{center}
{\bf\Large A Deformation of Quantum Dynamics}

\vspace{5pt}

{\bf\Large through the Phase Space Path Integral}\footnote{Contribution to the Proceedings of the Fifth International
Workshop on Contemporary Problems in Mathematical Physics, Cotonou, Republic of Benin, October 27--November 2, 2007,
eds. Jan Govaerts and M. Norbert Hounkonnou (International Chair in Mathematical Physics and Applications,
ICMPA-UNESCO, Cotonou, Republic of Benin, 2008), pp.~170--186.}
\end{center}
\vspace{10pt}
\begin{center}
Jan GOVAERTS$^{\dagger,\star,\ddagger}$
and Olivier MATTELAER$^{\dagger}$\\
\vspace{5pt}
$^\dagger${\sl Center for Particle Physics and Phenomenology (CP3),\\
Institut de Physique Nucl\'eaire, Universit\'e catholique de Louvain (U.C.L.),\\
2, Chemin du Cyclotron, B-1348 Louvain-la-Neuve, Belgium}\\
{\it E-Mail: Jan.Govaerts@uclouvain.be, Olivier.Mattelaer@uclouvain.be}\\
\vspace{7pt}
$^\star${\sl Fellow, Stellenbosch Institute for Advanced Study (STIAS),\\
7600 Stellenbosch, Republic of South Africa}\\
\vspace{7pt}
$^\ddagger${\sl International Chair in Mathematical Physics and Applications (ICMPA-UNESCO Chair),\\
University of Abomey--Calavi, 072 B. P. 50, Cotonou, Republic of Benin}
\end{center}

\vspace{15pt}

\begin{quote}
Using a regularised construction of the phase space path integral due to Ingrid Daubechies and John Klauder
which involves a time scale ultimately taken to vanish, and motivated by the general programme towards
a noncommutative space(time) geometry, physical consequences of assuming this time parameter to provide rather a new
fundamental time scale are explored in the context of the one dimensional harmonic oscillator. Some
tantalising results are achieved, which raise intriguing prospects when extrapolated
to the quantum field theory and gravitational contexts.
\end{quote}

\vspace{10pt}

\section{Introduction}
\label{Gov2.Sec1}

\subsection{Motivation}

From quite a number of vantage points, it could be argued that the phase space formulation
of a dynamical system is certainly at least as important, insightful and fundamental as its
configuration space formulation, if not far more. In other words, the extra dimensions associated to the momenta,
$p_i$, conjugate to some configuration space coordinates, $x^i$ $(i=1,2,\ldots,d)$, may well provide precisely
a new realm to be explored in the search for a final unification with extra degrees of freedom,
on a par with the idea of ordinary spacelike extra dimensions. Indeed, and this is especially true and relevant
when it comes to a gauge invariant dynamics with its first-class constraints and Hamiltonian,
the basic formulation\cite{Gov2.GovBook} of a dynamics is provided by the Hamiltonian first order action principle
defined over phase space, of the form
\begin{equation}
S[x^i,p_i]=\int dt\,\left[\dot{x}^i\,p_i\,-\,H(x^i,p_i)\,+\,\frac{dF(x^i,p_i)}{dt}\right],
\end{equation}
$H(x^i,p_i)$ being the Hamiltonian of the system which, through the (canonical) Poisson brackets
associated to the terms $\dot{x}^ip_i$ in the above expression, generates the time evolution of the
system (and $F(x^i,p_i)$ being an arbitrary function of which the Hamiltonian equations of motion remain
independent). In particular, it is in general possible to reduce the conjugate momenta, $p_i$, through the
Hamiltonian equations of motion for the configuration space variables, $x^i$, and recover the
configuration space action principle for the same system. However from quite a number of perspectives,
it proves important to consider the dynamics from its Hamiltonian formulation. In this respect, let us
only mention the appearance of so-called dynamical symmetries, which are symmetries of the Hamiltonian
dynamics of some systems and yet are not symmetries of their Lagrangian formulation.
Or at least as equally important, the fact that one of the royal paths into quantum physics is through
canonical operator quantisation which relies precisely on the Hamiltonian formulation of a dynamics,
and in particular on the Heisenberg commutation relations,
\begin{equation}
\left[\hat{x}^i,\hat{p}_j\right]=i\hbar\delta^i_j,\qquad i,j=1,2,\ldots,d,
\end{equation}
in direct correspondence with the classical Poisson brackets of these phase space canonical coordinates.
Within the context of gauge invariant dynamics it turns out that the full gauge content of such
dynamics is best realised through its Hamiltonian formulation since in actual fact, quite often a given Lagrangian
formulation of such a dynamics effects already, albeit implicitly, a partial gauge fixing\cite{Gov2.GovBook}.

The relation between quantisation and the phase space formulation of a dynamics also touches onto
a pressing issue nowadays in the search for a framework which would unify consistently a quantum description
of both the gravitational interaction and the three other known fundamental quantum interactions. Many arguments
and indications suggest that most probably such a quantum unification including gravity will have too rely,
or imply, a new type of geometry for spacetime, in which quantum effects would lead to quantum properties
for the basic coordinates of spacetime, namely such that the latter would become noncommuting operators
rather than being simply $c$-numbers as in ordinary differential geometry of the continuum. Indeed, such a final
unification would bring into a single arena of physics the fundamental three constants of mechanics and
gravity, namely the speed of light in vacuum, $c$, Planck's constant of quantum physics, $\hbar$, and
Newton's constant of gravity, $G_N$. Out of these three constants one may define absolute scales of
length, time, energy, mass and momentum, indeed purely spacetime and mechanical concepts. For instance,
one has
\begin{equation}
\ell_0=\sqrt{\hbar\frac{G_N}{c^3}},\qquad
p_0=\sqrt{\hbar\frac{c^3}{G_N}},\qquad
\tau_0=\sqrt{\hbar\frac{G_N}{c^5}},
\end{equation}
corresponding to fundamental scales of length, momentum and time, respectively,
known as the Planck scales.

Much work has thus already been devoted to attempts at formulating so-called noncommutative geometries.
Spurred on by what happens in quantum phase space as characterised by the above Heisenberg algebra for
the phase space coordinates as quantum operators, in the simplest setting deformations of ordinary
quantum mechanics of the following form have been considered,
\begin{equation}
\left[\hat{x}^i,\hat{x}^j\right]=i\hbar\theta^{ij},\qquad
\left[\hat{x}^i,\hat{p}_j\right]=i\hbar\delta^i_j,\qquad
\left[\hat{p}_i,\hat{p}_j\right]=i\hbar\lambda_{ij},
\label{eq:NCd}
\end{equation}
$\theta^{ij}$ and $\lambda_{ij}$ being some constant antisymmetric coefficients. More generally
in a Lorentz covariant context, a similar deformation would entail relations such as,
\begin{equation}
\left[\hat{x}^\mu,\hat{x}^\nu\right]=i\hbar\theta^{\mu\nu},\qquad
\left[\hat{x}^\mu,\hat{p}^\nu\right]=i\hbar\eta^{\mu\nu},\qquad
\left[\hat{p}^\mu,\hat{p}^\nu\right]=i\hbar\lambda^{\mu\nu},
\end{equation}
where this time, in addition to the constant antisymmetric coefficients $\theta^{\mu\nu}$ and
$\lambda^{\mu\nu}$, $\eta^{\mu\nu}$ stands for the Minkowski spacetime metric. In particular
within the framework of string and M-theory, in the presence of background fields and in specific
limits there appear precisely such types of noncommutative spacetime coordinates.

As a matter of illustration, taking seriously the point of view that all phase space coordinates should
be considered on equal terms, for the sake of the discussion let us restrict to a four dimensional (euclidean)
phase space (a two dimensional configuration space) and measure all phase space coordinates in units of
some length scale, $\ell_0$, or momentum scale, $p_0$, while time is to be measured in units of some time
scale, $\tau_0$ (none of these scales need at this stage to coincide with the above Planck scales).
The simplest deformation of the ordinary Heisenberg
algebra would thus read
\begin{equation}
\left[\frac{\hat{x}^i}{\ell_0},\frac{\hat{x}^j}{\ell_0}\right]=i\epsilon^{ij},\qquad
\left[\frac{\hat{x}^i}{\ell_0},\frac{\hat{p}_j}{p_0}\right]=i\delta^i_j,\qquad
\left[\frac{\hat{p}_i}{p_0},\frac{\hat{p}_j}{p_0}\right]=i\epsilon_{ij},\qquad i,j=1,2,
\end{equation}
or equivalently,
\begin{equation}
\left[\hat{x}^i,\hat{x}^j\right]=i\hbar\,\epsilon^{ij}\,\left(\frac{\ell_0}{p_0}\right),\qquad
\left[\hat{x}^i,\hat{p}_j\right]=i\hbar,\qquad
\left[\hat{p}_i,\hat{p}_j\right]=i\hbar\,\epsilon_{ij}\,\left(\frac{p_0}{\ell_0}\right),\qquad i,j=1,2,
\label{eq:NC2}
\end{equation}
in which the following identification has been made,
\begin{equation}
\ell_0\,p_0=\hbar.
\end{equation}
Incidentally, by an appropriate linear redefinition of the operators $\hat{p}_i$ involving the coordinate
operators $\hat{x}^i$, it is always possible to transform the above algebra to one in which the momentum-momentum
commutators vanish identically (and likewise for the position-position ones through an appropriate redefinition
of the coordinate operators involving the momentum ones).

When using for the fundamental scales $\ell_0$, $p_0$ and $\tau_0$ the Planck scales specified above,
one obtains some intriguing relationships,
\begin{equation}
\left[\hat{x}^i,\hat{x}^j\right]=i\hbar\,\epsilon^{ij}\,\left(\frac{G_N}{c^3}\right),\qquad
\left[\hat{x}^i,\hat{p}_j\right]=i\hbar\,\delta^i_j,\qquad
\left[\hat{p}_i,\hat{p}_j\right]=i\hbar\,\epsilon_{ij}\,\left(\frac{c^3}{G_N}\right),\qquad i,j=1,2.
\end{equation}
Since the non vanishing momentum-momentum commutator may be gauged away through an appropriate linear
redefinition of the momentum operators, let us concentrate on the first commutator, also because of the general
perspective of a possible noncommutative geometry in configuration space\footnote{A similar argument could be made
for the momentum-momentum commutators, but then with a behaviour opposite to the one to be described.}.
It is quite noteworthy that in the limit of a vanishing gravitational strength, the space-space commutators are
then vanishing, implying a ordinary commutative configuration space. Or likewise in the nonrelativistic limit
with $G_N/c^3\rightarrow 0$, a commutative configuration space is once again recovered. It is only if both a relativistic and
a gravitational framework is considered that a noncommutative structure of configuration space may possibly emerge,
on dimensional grounds. Note also that the limit $G_N/c^3\rightarrow 0$ makes the original momentum operators
$\hat{p}_i$ to become singular, while nonetheless the configuration space dependent twisted commuting momentum
operators remain well defined at all stages. In a certain sense thus, turning on the quantity $G_N/c^3$, namely
relativistic quantum gravity, brings to the fore the hidden noncommutative configuration space dimensions of phase space.

\subsection{Illustration}

To make the last comment more explicit, let us consider a system in which extra phase space dimensions emerge
through similar limits while at the same time leading to noncommutative structures in configuration space.
The simplest such setting is that of the celebrated Landau problem, describing the motion of a charged particle
confined to a plane and submitted to a static and homogeneous magnetic field perpendicular to that plane.
The associated dynamics follows from the Lagrange function, expressed in cartesian coordinates and in
the symmetric gauge for the magnetic field,
\begin{equation}
L=\frac{1}{2}m\left(\dot{x}^2+\dot{y}^2\right)+\frac{1}{2}B\left(\dot{x}y-x\dot{y}\right)-V(x,y).
\end{equation}
Here, an extra interaction potential energy, $V(x,y)$, has been included, while the particle's electric charge
has been absorbed in the definition of the magnetic field $B$. Denoting as $p_x$ and $p_y$ the
momenta canonically conjugate to the coordinates $x$ and $y$, respectively, the Hamiltonian generating the
time dependence of the trajectories of the system in this four dimensional phase space is
\begin{equation}
H=\frac{1}{2m}\left(p_x-\frac{1}{2}By\right)^2+\frac{1}{2m}\left(p_y+\frac{1}{2}Bx\right)^2+V(x,y).
\end{equation}

By analogy with the above limit, $\ell_0/p_0\rightarrow 0$ or $G_N/c^3\rightarrow 0$, let us now
consider the limit of the above Landau model in which $m\rightarrow 0$. In order to retain configurations
of finite energy only, one must enforce in this limit the following two constraints,
\begin{equation}
\phi_1=p_x-\frac{1}{2}By=0,\qquad
\phi_2=p_y+\frac{1}{2}Bx=0.
\end{equation}
Considering their Poisson bracket, $\{\phi_1,\phi_2\}=-B$, it turns out they define second-class
constraints\cite{Gov2.GovBook}. Solving the latter through the construction of the associated Dirac
brackets, one finds
\begin{equation}
\{x,y\}_{\rm Dirac}=\frac{1}{B}.
\end{equation}
In other words, through the limit $m\rightarrow 0$, the initially four dimensional phase space has reduced
to a two dimensional phase space which, as a matter of fact, corresponds to the two dimensional configuration space
of the original system. Or conversely, by turning on the parameter $m$, out of a two dimensional phase space
there may emerge a four dimensional one. Furthermore at the quantum level, through that limit $m\rightarrow 0$,
the initially commuting configuration space operators become noncommuting, providing the simplest instance
of an example of a noncommutative geometry,
\begin{equation}
\left[\hat{x},\hat{y}\right]=0,\qquad
\left[\hat{x},\hat{y}\right]_{\rm Dirac}=i\hbar\frac{1}{B}.
\end{equation}
As a matter of fact, when considered within the quantised original system, the limit $m\rightarrow 0$ effects
the projection of the quantum Landau problem onto the lowest Landau level, in presence of the magnetic field,
corresponding to a limit in which the energy gap between Landau levels set by the cyclotron frequency $|B/m|$
grows infinite.

From the dynamical perspective, in the same limit the original Lagrange function of the system reduces to the action,
\begin{equation}
S_{m=0}[x,y]=\int\,dt\left[\frac{1}{2}B\left(\dot{x}y-x\dot{y}\right)-V(x,y)\right].
\end{equation}
Being linear in the first order time derivatives of the remaining degrees of freedom $x$ and $y$, this is indeed the
phase space action for a system of the form
\begin{equation}
S[q,p]=\int\,dt\left[\frac{1}{2}\left(\dot{q}p-q\dot{p}\right)-H(q,p)\right],
\end{equation}
with the correspondences,
\begin{equation}
x\longleftrightarrow q,\qquad
By\longleftrightarrow p,\qquad
V(x,y)\longleftrightarrow H(q,p),\qquad
\left\{x,By\right\}_{\rm Dirac}=1\longleftrightarrow \{q,p\}=1.
\end{equation}
The original Lagrangian dynamics on the $(x,y)$ configuration space corresponding to a system with
two degrees of freedom has reduced to the Hamiltonian dynamics on the $(x,y)$ phase space corresponding to
a single degree of freedom system. Conversely, turning on the mass parameter $m$, out of the one dimensional
system with the two dimensional phase space of canonically conjugate variables $(x,By)$ and Hamiltonian $V(x,y)$
emerges a system with a two dimensional configuration space $(x,y)$ and interaction potential energy $V(x,y)$ subjected
to the homogeneous magnetic field $B$, which in effect realises the symplectic structure of the system
in the limit $m\rightarrow 0$. To the latter system is thus associated a four dimensional phase space, $(x,p_x;y,p_y)$.

As a specific illustration consider the two dimensional particle subjected not only to the magnetic field but also
a spherically symmetric harmonic well,
\begin{equation}
L=\frac{1}{2}m\left(\dot{x}^2+\dot{y}^2\right)+\frac{1}{2}B\left(\dot{x}y-x\dot{y}\right)
-\frac{1}{2}k\left(x^2+y^2\right),\qquad k>0.
\end{equation}
In the limit $m\rightarrow 0$, this system reduces to the Hamiltonian one,
\begin{equation}
L_{m=0}=\frac{1}{2}\left(\dot{x}p-x\dot{p}\right)-\frac{1}{2m_0}p^2-\frac{1}{2}k_0x^2,
\end{equation}
in terms of quantities rescaled in an obvious manner. The latter expression corresponds to the Hamiltonian
phase space formulation of the one dimensional harmonic oscillator, of which the classical trajectories are
simply ellipses in the $(x,p)$ plane along which the system evolves with a specific orientation. When the
parameter $m$ is turned on again, these ellipses get fuzzied or blurred out by having the system's trajectoires
having now two contributions, one being still that of the oriented ellipse on top of which is superposed
a circular motion with the opposite orientation and possessing a periodicity set by the cyclotron frequency,
as well as a curvature radius which grows ever tighter and smaller as the mass parameter $m$ approaches a vanishing value.
The ellipse is the guiding trajectory for the magnetic center of the particle in the plane, around which it effects
a circular motion with the opposite orientation and which is induced by the magnetic field.

It is tantalising to also speculate about the fact that when turning on the
parameter $m$, a trajectory which initially is perfectly well defined and sharp, becomes blurred and fuzzy
about a guiding center, a feature which is very much reminiscent of the impossibility in quantum mechanics of
measuring with perfect precision both at the same time two quantities that do not commute. Could Heisenberg's
uncertainty relations of quantum mechanics be related to some hitherto unidentified fundamental constant of physics
akin to the above parameter $m$, which, albeit extremely small and possibly connected to Planck's scales, could
reveal emergent extra dimensions in phase space? Could it be that the phase spaces of our fundamental theories 
with their noncommutative algebras are rather the configuration spaces of some underlying dynamics taking
place in a still higher dimensional phase space? Could it be that the idea of extra dimensions of space should
in fact apply to phase space dimensions rather than simply configuration space? It is the purpose of the present
contribution to explore such a possibility, beginning with the simplest of non trivial dynamics, that of the one dimensional
harmonic oscillator, which is also what suffices to understand quantum field theories albeit in a perturbative regime.

\subsection{Unsatisfactory issues}

Taken at face value, the commutation relations (\ref{eq:NCd}) and (\ref{eq:NC2}) raise the following issue.
As is well known within Hamiltonian dynamics, Darboux's theorem assures us that there always exists a system of
coordinates on phase space such that the Poisson brackets take their canonical form. At the quantum operator
level and in the present instance, this means simply that there exist specific linear redefinitions of the
operators $\hat{x}^i$ and $\hat{p}_i$ leading to new operators $\hat{X}^i$ and $\hat{\Pi}_i$ defining the
ordinary Heisenberg algebra,
\begin{equation}
\left[\hat{X}^i,\hat{X}^j\right]=0,\qquad
\left[\hat{X}^i,\hat{\Pi}_j\right]=i\hbar\delta^i_j,\qquad
\left[\hat{\Pi}_i,\hat{\Pi}_j\right]=0.
\end{equation}
In other words, there is no intrinsic, invariant, covariant, or coordinate-free meaning to the statement that some
configuration space is noncommutative, since within the phase space context there always exists a
choice of canonical coordinate operators which are commutative. It is only provided some extraneous physical
information is imposed on the description of the system, which in effect would amount to requiring
that according to some reason the original coordinates $\hat{x}^i$ are to be considered to define
in an absolute way ``the" configuration space of the system, that a noncommutative configuration space
geometry could be given a specific meaning. Yet, such a situation runs counter to all
principles of a covariant formulation of the laws of physics, in particular that dynamics should remain
coordinate-free and invariant under arbitrary canonical transformations of phase space.

Of course, this situation is not new. In 1931 already, P. Dirac in his famous book on quantum physics\cite{Gov2.Dirac}
pointed out that the Heisenberg algebra and its representations, as such, are valid only provided the
coordinates $(x^i,p_i)$ of phase space are in fact cartesian coordinates for some implicit Euclidean metric
defined over phase space. This very feature also translates into the fact that, in spite of formal appearances,
Feynman's path integral over phase space is not invariant under canonical transformations of phase space.

It is precisely this situation with regards to parametrisations of phase space, quantisation and coordinate invariance
of the physical representation, which led John Klauder to suggest\cite{Gov2.Klauder1} that besides the symplectic structure
with which phase space comes equipped enabling thus the programme of canonical operator quantisation, one ought also to introduce
a compatible Riemannian metric over phase space, in whose absence it is impossible to give an invariant
physical meaning to a quantisation procedure. As a matter of fact this implicit or ``shadow metric" structure of phase space
even determines the quantum physical content of the system, and yet in the classical limit plays no r\^ole any longer in the
classical dynamics. The ensuing construction of the quantum path integral over phase space is such that it is now
invariant under any canonical transformations of phase space, giving it thus an intrinsic, invariant and coordinate-free
meaning.

The construction of this phase space path integral\cite{Gov2.Klauder1,Gov2.Klauder2,Gov2.Klauder3,Gov2.Klauder4,Gov2.Klauder5},
which relies on prior work by Ingrid Daubechies and John Klauder\cite{Gov2.KD1,Gov2.KD2,Gov2.KD3,Gov2.KD4,Gov2.KD5}, may be
interpreted as follows. In effect the Riemannian metric structure introduced on phase space and compatible with its
symplectic structure is brought into the quantum arena as an extra contribution to the functional integral measure
in the path integral. This extra factor corresponds to a Brownian motion component of the particle's motion over phase space\footnote{The
reason for this feature is that in actual fact the path integral construction is based on techniques of stochastic calculus.}
to which a specific diffusion time scale is associated, say $\tau_0$, which takes the form of a real exponential Gaussian factor
involving the invariant phase space line element constructed out of the Riemannian metric. Hence, the ordinary phase
space trajectories of the particle play now the r\^ole of guiding centers for trajectories which diffuse out of these centers,
leading to some fuzzied structure surrounding the guiding center whose time scale is set by the parameter $\tau_0$. And indeed all
ordinary quantum physical properties and results are recovered in the limit $\tau_0\rightarrow 0$\cite{Gov2.Klauder1}.

In other words, the parameter $\tau_0$ plays a r\^ole akin to that of the mass parameter, $m$, in the previous discussion.
The phase space of the original system becomes the configuration space of the $\tau_0$-extended system, of which the
Hamiltonian formulation involves a phase space whose dimension is twice that of the original system. Furthermore
the noncommutative quantum Heisenberg algebra of the original phase space degrees of freedom results from the limit $\tau_0\rightarrow 0$
of the extended quantised system, in which the original phase space degrees of freedom remain commutative. Hence, when keeping
the value of $\tau_0$ finite, extra phase space dimensions emerge from the original phase space and one has a particular type
of deformation of ordinary quantum physics related to the constant $\tau_0$. Such a deformation should result in a deformation
of the Heisenberg algebra of the original system, namely a certain form of noncommutative geometry, but which this time is given
an intrinsic, geometric invariant and coordinate-free meaning, and such that in the limit $\tau_0\rightarrow 0$ ordinary
quantum mechanics is recovered in the original phase space. The proposal of the present contribution is thus to explore
the eventual physical consequences of considering that the scale $\tau_0$ is possibly related to a new fundamental constant
in physics, in addition to $\hbar$ and $c$, leaving aside at this stage the issue of a connection with the gravitational
constant $G_N$. In any case, given all the successes of ordinary quantum physics the time scale $\tau_0$ ought to be
extremely small, presumably on the order of the Planck time scale, some $10^{-42}$ s.

Note however that the above scheme differs in a crucial aspect from the previous discussion in terms of the parameter $m$.
In the latter case the term linear in $m$ contributing to the action implies a pure imaginary phase factor multiplying
the path integral measure. However for what concerns the time scale $\tau_0$ and the associated Brownian motion Gaussian
factor, the latter being pure real exponential implies that the analogy between the two situations corresponds to a pure
positive imaginary mass parameter $m$. In that respect, Klauder's path integral construction adds onto the ordinary
probability amplitude properties of quantum mechanics an extra feature which in actual fact is that of a deterministic
statistical behaviour. The interplay between these two general origins for indeterminism in physics is bound to imply
interesting properties. And yet in the limit of a vanishing time scale $\tau_0$ the ordinary rules of quantum
indeterminism become solely relevant again. Such features are also reminiscent of Gerard~'t~Hooft's attempts at constructing
a deterministic quantum physics\cite{Gov2.tHooft}, eventually within a gravitational context.

\section{A Construction of the Phase Space Path Integral}
\label{Gov2.Sec2}

In order to describe the construction of the phase space path integral discussed in Ref.\cite{Gov2.Klauder1},
which is the culmination of the work in
Refs.\cite{Gov2.Klauder1,Gov2.Klauder2,Gov2.Klauder3,Gov2.Klauder4,Gov2.Klauder5,Gov2.KD1,Gov2.KD2,Gov2.KD3,Gov2.KD4,Gov2.KD5},
let us restrict to a single degree of freedom system, and work in natural units such that $\hbar=1$ while
all other possible independent dimensionful parameters are also set to unity. Within the phase space formulation at the
quantum level, such a dynamics is then characterised\cite{Gov2.Gov1} by operators $\hat{q}$ and $\hat{p}$ obeying the Heisenberg algebra,
\begin{equation}
\left[\hat{q},\hat{p}\right]=i\mathbb{I},\qquad
\hat{q}^\dagger=\hat{q},\qquad
\hat{p}^\dagger=\hat{p},
\end{equation}
of which the time evolution is governed by some quantum Hamiltonian $\hat{H}(\hat{q},\hat{p})$.

Besides the usual configuration and momentum space representations of the Heisenberg algebra constructed out
of the bases of position and momentum eigenstates, respectively,
\begin{equation}
\hat{q}|q\rangle=q|q\rangle,\qquad
\hat{p}|p\rangle=p|p\rangle,\qquad q,p\in\mathbb{R},
\end{equation}
introducing the Fock algebra operators,
\begin{equation}
a=\frac{1}{\sqrt{2}}\left(\hat{q}+i\hat{p}\right),\qquad
a^\dagger=\frac{1}{\sqrt{2}}\left(\hat{q}-i\hat{p}\right),\qquad
\left[a,a^\dagger\right]=\mathbb{I},
\end{equation}
yet another orthonormalised basis of the representing Hilbert space is provided by the Fock states $|n\rangle$, $n\in\mathbb{N}$, such that
\begin{equation}
\langle n|\ell\rangle=\delta_{n,\ell},\qquad
|n\rangle=\frac{1}{\sqrt{n!}}\left(a^\dagger\right)^n|\Omega_0\rangle,\qquad
a|\Omega_0\rangle=0,\qquad \langle \Omega_0|\Omega_0\rangle=1.
\end{equation}

Based on these Fock states, coherent states, $|q,p\rangle$, in that Hilbert space in one-to-one correspondence with the classical
phase space states labelled by the pair $(q,p)$ are then constructed as\cite{Gov2.KS},
\begin{equation}
|z\rangle\equiv |q,p\rangle=e^{-i\left(q\hat{p}-p\hat{q}\right)}|\Omega_0\rangle=
e^{-\frac{1}{2}|z|^2}\,e^{z a^\dagger}\,|\Omega_0\rangle,
\end{equation}
with the correspondence
\begin{equation}
z=\frac{1}{\sqrt{2}}\left(q+ip\right).
\end{equation}
These coherent states possess a series of quite remarkable properties\cite{Gov2.KS}. Even though they generate the whole
Hilbert space, namely one has the following decomposition of the unit operator,
\begin{equation}
\int_{(\infty)}\frac{dq dp}{2\pi}\,|q,p\rangle\langle q,p|=\mathbb{I},
\end{equation}
these normalised states provide an over-complete basis since they are not orthogonal,
\begin{equation}
\langle z_2|z_1\rangle=\langle q_2,p_2|q_1,p_1\rangle=
e^{-\frac{1}{2}|z_2|^2-\frac{1}{2}|z_1|^2+\bar{z}_2 z_1}=
e^{-\frac{1}{4}\left((q_2-q_1)^2+(p_2-p_1)^2\right)+
\frac{1}{2}i(q_2p_1-q_1p_2)}.
\end{equation}
Furthermore these coherent states define sharp quantum states in phase space, by which is meant that one has
\begin{equation}
\langle q,p|\hat{q}|q,p\rangle=q,\qquad
\langle q,p|\hat{p}|q,p\rangle=p,
\end{equation}
while they also saturate the Heisenberg uncertainty relations,
\begin{equation}
\Delta q\,\Delta p=\frac{1}{2}.
\end{equation}
Such quantum states are thus the closest possible to being actual classical states. Finally, any bounded
operator, $\hat{A}$, acting on that Hilbert space possesses a diagonal kernel representation in terms of these
coherent states,
\begin{equation}
\hat{A}=\int_{(\infty)}\frac{dq dp}{2\pi}\,|q,p\rangle\,a(q,p)\,\langle q,p|,\qquad
a(q,p)=e^{-\frac{1}{2}\left(\frac{\partial^2}{\partial q^2}+\frac{\partial^2}{\partial p^2}\right)}\,
\langle q,p|\hat{A}|q,p\rangle .
\end{equation}
Thus in particular the Hamiltonian operator $\hat{H}(\hat{q},\hat{p})$ itself is represented by the kernel
\begin{equation}
h(q,p)=e^{-\frac{1}{2}\left(\frac{\partial^2}{\partial q^2}+\frac{\partial^2}{\partial p^2}\right)}\,
\langle q,p|\hat{H}(\hat{q},\hat{p})|q,p\rangle,\qquad
\hat{H}=\int_{(\infty)}\frac{dq dp}{2\pi}\,|q,p\rangle\,h(q,p)\,\langle q,p|.
\label{eq:ck}
\end{equation}
Note that in the case of a Hamiltonian which is purely quadratic in both $\hat{q}^2$ and $\hat{p}^2$,
as is that for the harmonic oscillator, the kernel $h(q,p)$ differs from the classical Hamiltonian
$H(q,p)$ by a constant term only, corresponding simply to the quantum vacuum energy.

Using the representation (\ref{eq:ck}) it then becomes possible to set up a path integral representation
over phase space of the quantum evolution operator associated to the Hamiltonian operator $\hat{H}$,
\begin{equation}
\hat{U}_0(t_2,t_1)=e^{-iT_{21}\,\hat{H}},\qquad T_{21}=t_2-t_1.
\end{equation}
Introducing an equally spaced time slicing of the time interval, $T_{21}=t_2-t_1$, in $N$ integer steps with spacing
\begin{equation}
\epsilon=\frac{T_{21}}{N}=\frac{t_2-t_1}{N},\qquad N\in\mathbb{N}^*,
\end{equation}
and writing then
\begin{equation}
\hat{U}_0(t_2,t_1)=\left(e^{-i\epsilon\hat{H}}\right)^N=
\lim_{N\rightarrow \infty}\left(\mathbb{I}-i\epsilon\hat{H}\right)^N,
\end{equation}
one may now use the following kernel representation for each of the $N$ operator factors
contributing in the above representation of the unitary quantum evolution operator $\hat{U}_0(t_2,t_1)$,
\begin{equation}
\mathbb{I}-i\epsilon\hat{H}=\int_{(\infty)}\frac{dq dp}{2\pi}
|q,p\rangle\,\Big(1-i\epsilon h(q,p)\Big)\,\langle q,p|.
\end{equation}
Consequently one then finds,
\begin{eqnarray}
\langle q_2,p_2|\hat{U}_0(t_2,t_1)|q_1,p_1\rangle &=& \lim_{N\rightarrow\infty}
\int_{(\infty)}\prod_{\alpha=1}^{N}\frac{dq_\alpha dp_\alpha}{2\pi}\times \nonumber \\
\times \exp i\sum_{\alpha=0}^{N}\epsilon &\Big[&\frac{1}{2}\left(\frac{q_{\alpha+1}-q_\alpha}{\epsilon}p_\alpha\,-\,
q_\alpha\frac{p_{\alpha+1}-p_\alpha}{\epsilon}\right)\,-\,h(q_\alpha,p_\alpha)\,+  \nonumber \\
&+&  \frac{1}{4}i\epsilon
\left(\left(\frac{q_{\alpha+1}-q_\alpha}{\epsilon}\right)^2+\left(\frac{p_{\alpha+1}-p_\alpha}{\epsilon}\right)^2\right)\Big],
\label{eq:PSCSPI}
\end{eqnarray}
with of course $(q,p)_{\alpha=0}=(q_1,p_1)$ and $(q,p)_{\alpha=N+1}=(q_2,p_2)$.
This expression provides an exact integral representation for the given phase space coherent state matrix elements of the
quantum evolution operator. Usually in the formal limit $\epsilon\rightarrow 0$, the contribution of the very last
line of the above representation is ignored, leading to the following formal expression for the phase space
coherent state functional integral,
\begin{equation}
\langle q_2,p_2|\hat{U}_0(t_2,t_1)|q_1,p_1\rangle = \int_{(q_1,p_1)}^{(q_2,p_2)}\left[\frac{{\cal D}q{\cal D}p}{2\pi}\right]\,
e^{i\int_{t_1}^{t_2}dt\left[\frac{1}{2}\left(\dot{q}p-q\dot{p}\right)-h(q,p)\right]},
\end{equation}
thus in terms of the Hamiltonian first-order action of the system.
However the factor which is being ignored being real exponential Gaussian, is crucial in controlling the convergence
properties of the functional integral. To keep it explicit, let us introduce a finite time scale $\tau_0>0$ in place
of the factor $\epsilon/2$ appearing in the last line of (\ref{eq:PSCSPI}), and then let it run to zero only
after having evaluated the path integral,
\begin{eqnarray}
\langle q_2,p_2|\hat{U}_0(t_2,t_1)|q_1,p_1\rangle &=& \lim_{\tau_0\rightarrow 0}
\int_{(q_1,p_1)}^{(q_2,p_2)}\left[\frac{{\cal D}q{\cal D}p}{2\pi}\right]\,
e^{i\int_{t_1}^{t_2}dt\left[\frac{1}{2}\left(\dot{q}p-q\dot{p}\right)-h(q,p)+\frac{1}{2}i\tau_0
\left(\dot{q}^2+\dot{p}^2\right)\right]} \nonumber \\
&=& \lim_{\tau_0\rightarrow 0} \int_{(q_1,p_1)}^{(q_2,p_2)}\left[\frac{{\cal D}q{\cal D}p}{2\pi}\right]\,
e^{-\int_{t_1}^{t_2}dt\,\frac{1}{2}\tau_0\left(\dot{q}^2+\dot{p}^2\right)}\,
e^{i\int_{t_1}^{t_2}dt\left[\frac{1}{2}\left(\dot{q}p-q\dot{p}\right)-h(q,p)\right]}.
\label{eq:PI1}
\end{eqnarray}
In this manner an effective Landau problem arises for a particle with a pure positive imaginary mass $i\tau_0/2$ moving
in the plane $(q,p)$ now as its configuration space. The phase space of the new extended system is thus four dimensional,
the nonvanishing constant $\tau_0$ having thus made manifest two extra phase space dimensions emergent from the
original two dimensional phase space $(q,p)$. The limit $\tau_0\rightarrow 0$ is analogous to taking the
lowest Landau level projection of the extended system, such that the dynamics becomes confined to the lowest Landau level
which coincides with the Hilbert space of the original one dimensional system.

This provides an heuristic argument for the construction of the phase space coherent state path integral
by Ingrid Daubechies and John Klauder based on the stochastic calculus of quantum trajectoires in phase space.
The analysis by these authors, culminating in Ref.\cite{Gov2.Klauder1}, leads to a definition of the
phase space functional integral measure which is that of the Wiener measure, in effect in physics terms
the measure for statistical Brownian motion in phase space, once a Riemannian metric is defined over phase
space. Indeed, in the above expression, the extra Gaussian factor related to $\tau_0$ is such a measure
factor associated to the Euclidean metric over phase space. Finally, the Wiener measure needs to be normalised
in such a manner that the limit $\tau_0\rightarrow 0$ leads to a finite result, which requires a specific
subtraction of the vacuum quantum energy such that the extended system projects to the lowest Landau level
states of finite energy in that limit.

The noteworthy results discussed in Ref.\cite{Gov2.Klauder1} are the following. First, that for large classes
of Hamiltonian functions $H(q,p)$\cite{Gov2.KD1,Gov2.KD1.2}, in the limit $\tau_0\rightarrow 0$ the result for the path integral
on the r.h.s. of (\ref{eq:PI1}) for $T_{21}>0$ indeed reduces to the matrix element on the l.h.s., namely that of the ordinary quantum physics
of the system, and is thus well-defined. Second, that for $T_{21}>0$ the path integral on the r.h.s. is well-defined for
any finite value of $\tau_0$. Third, that the resulting formulation of quantum amplitudes is invariant under
canonical transformations in phase space. Hence this is indeed the formalism which is required in order to
construct a coordinate-free realisation of a noncommutative geometry over phase space, and in particular
in configuration space.

Incidentally, note that the introduction of the Brownian motion Gaussian factor in the integration measure
of the path integral effects a regularisation which is quite analogous to the following regularisation of a pure
imaginary Gaussian integral\cite{Gov2.Klauder1},
\begin{equation}
\int_0^\infty dy\,e^{i\alpha y^2}=\lim_{\tau_0\rightarrow 0^+}\int_0^\infty dy\,e^{i\alpha y^2-\frac{1}{2}\tau_0 y^2},
\end{equation}
namely by analytic continuation in the complex $\alpha$ plane.

Keeping the value of $\tau_0$ finite, the above construction of the phase space coherent state path integral for
a given system thus defines a specific deformation of ordinary quantum dynamics. Taking the point of view that
a scale such as $\tau_0$ could provide for a new fundamental constant of physics associated to a form of
noncommutative geometry is certainly worth exploring, which is the purpose of the present contribution. The ordinary
path integral is ill-defined, while a finite $\tau_0$ provides a regularisation for it. Could it be that there is
more to the scale $\tau_0$ than simply a regularisation of our present day physical representations at the smallest (time, hence
space) scales? As a first step into that direction, this deformation of quantum dynamics is considered hereafter
in the context of the one dimensional harmonic oscillator.

\section{The Ordinary Harmonic Oscillator}
\label{Gov2.Sec3}

The classical dynamics of the one dimensional harmonic oscillator follows from the action principle
\begin{equation}
S[x]=\int dt\,\left[\frac{1}{2}m\dot{x}^2-\frac{1}{2}m\omega^2 x^2\right],
\end{equation}
$\omega>0$ being the angular frequency of the oscillator and $m$ its mass. Its Hamiltonian formulation
with momentum conjugate $p$ is characterised by the Hamilton function
\begin{equation}
H(x,p)=\frac{1}{2m}p^2+\frac{1}{2}m\omega^2 x^2.
\end{equation}
In particular, given boundary values $x_2=x(t_2)$ and $x_1=x(t_1)$ for the time interval $t_1\le t\le t_2$
in configuration space, the value of the classical action given the associated classical trajectory is
\begin{equation}
S_c=\frac{m\omega^2}{2\sin\omega T_{21}}\left[\left(x^2_2+x^2_1\right)\cos\omega T_{21}\,-\,
2 x_2x_1\right],\qquad T_{21}=t_2-t_1.
\label{eq:Sc}
\end{equation}

At the quantum level, the Fock space generators are defined as
\begin{equation}
a=\sqrt{\frac{m\omega}{2\hbar}}\left(\hat{x}+\frac{i}{m\omega}\hat{p}\right),\qquad
a^\dagger=\sqrt{\frac{m\omega}{2\hbar}}\left(\hat{x}-\frac{i}{m\omega}\hat{p}\right),
\end{equation}
which are such that $\left[a,a^\dagger\right]=\mathbb{I}$. Consequently the Hamiltonian,
\begin{equation}
\hat{H}=\frac{1}{2m}\hat{p}^2+\frac{1}{2}m\omega^2\hat{q}^2=\hbar\omega\left(a^\dagger a+\frac{1}{2}\right),
\end{equation}
is diagonalised in the Fock basis $|n\rangle$ $(n\in\mathbb{N})$, with
\begin{equation}
\hat{H}|n\rangle=E_n|n\rangle,\qquad E_n=\hbar\omega(n+1),\qquad n\in\mathbb{N}.
\end{equation}
The corresponding phase space coherent states are thus given as
\begin{equation}
|z\rangle=|x,p\rangle=e^{-\frac{i}{\hbar}\left(x\hat{p}-p\hat{x}\right)}\,|\Omega_0\rangle=
e^{-\frac{1}{2}|z|^2}\,e^{z a^\dagger}\,|\Omega_0\rangle,\qquad
z=\sqrt{\frac{m\omega}{2\hbar}}\left(x+\frac{i}{m\omega}p\right).
\end{equation}

Considering the quantum evolution operator, $\hat{U}_0(t_2,t_1)=e^{-\frac{i}{\hbar}(t_2-t_1)\hat{H}}$,
it is possible to obtain its configuration space matrix elements in the form
\begin{equation}
\langle x_2|\hat{U}_0(t_2,t_1)|x_1\rangle=
\left(\frac{m}{2i\pi\hbar T_{21}}\,\frac{\omega T_{21}}{\sin\omega T_{21}}\right)^{1/2}\,
e^{\frac{i}{\hbar}S_c},
\end{equation}
$S_c$ being the value of the classical action for the given boundary values, see (\ref{eq:Sc}).
Note that one has the limit
\begin{equation}
\lim_{T_{21}\rightarrow 0}\langle x_2|\hat{U}_0(t_2,t_1)|x_1\rangle=\delta(x_2-x_1),
\end{equation}
as it should of course, given the normalisation of the configuration space eigenstates, $\langle x|x'\rangle=\delta(x-x')$.

In a likewise manner it is possible to compute the phase space coherent state matrix elements of
the quantum evolution operator,
\begin{equation}
K_0(x_2,p_2,t_2;x_1,p_1,t_1)=\langle x_2,p_2|\hat{U}_0(t_2,t_1)|x_1,p_1\rangle.
\end{equation}
One finds,
\begin{eqnarray}
K_0(x_2,p_2,t_2;x_1,p_1,t_1) &=& \ \ \ e^{-\frac{1}{2}i\omega T_{21}}\,
e^{\frac{i}{2\hbar}\left(x_2p_1-x_1p_2\right) e^{-i\omega T_{21}}}\times \nonumber \\
&&\times\ e^{-\frac{m\omega}{4\hbar}\left(x^2_2+x^2_1-2x_2x_1e^{-i\omega T_{21}}\right)\,-\,
\frac{1}{4\hbar}\frac{1}{m\omega}\left(p^2_2+p^2_1-2p_2p_1e^{-i\omega T_{21}}\right)},
\end{eqnarray}
with in particular
\begin{equation}
\lim_{T_{21}\rightarrow 0} K_0(x_2,p_2,t_2;x_1,p_1,t_1)=
e^{\frac{i}{2\hbar}\left(x_2p_1-x_1p_2\right)}\,
e^{-\frac{m\omega}{4\hbar}\left(x_2-x_1\right)^2
-\frac{1}{4\hbar}\frac{1}{m\omega}\left(p_2-p_1\right)^2}=
\langle x_2,p_2|x_1,p_1\rangle.
\end{equation}

Note that the kernel $K_0(x_2,p_2,t_2;x_1,p_1,t_1)$ also obeys the convolution or reproducing property,
\begin{equation}
\int_{(\infty)}\frac{dx_2 dp_2}{2\pi\hbar}\,
K_0(x_3,p_3,t_3;x_2,p_2,t_2)\,K_0(x_2,p_2,t_2;x_1,p_1,t_1)=
K_0(x_3,p_3,t_3;x_1,p_1,t_1),
\label{eq:repro1}
\end{equation}
as well as the unitary property,
\begin{equation}
K^*_0(x_2,p_2,t_2;x_1,p_1,t_1)=K_0(x_1,p_1,t_1;x_2,p_2,t_2),
\end{equation}
as it should since as an abstract operator the evolution operator $\hat{U}_0(t_2,t_1)$ obeys precisely
these same two properties,
\begin{equation}
\hat{U}_0(t_3,t_2)\,\hat{U}_0(t_2,t_1)=\hat{U}_0(t_3,t_1),\qquad
\hat{U}^\dagger_0(t_2,t_1)=\hat{U}_0(t_1,t_2).
\end{equation}
In particular, the reproducing property (\ref{eq:repro1}) for the evolution kernel $K_0(x_2,p_2,t_2;x_1,p_1,t_1)$
directly follows from the latter operator convolution property through a simple application of the overcompleteness
relation for the phase space coherent states,
\begin{equation}
\int_{(\infty)}\frac{dx dp}{2\pi\hbar}\,|x,p\rangle\langle x,p|=\mathbb{I}.
\end{equation}

\section{The Deformed Harmonic Oscillator}
\label{Gov2.Sec4}

\subsection{Setting the stage}

Given our previous discussion and considerations, let us now consider the kernel defined by
the deformed phase space path integral including the finite time scale $\tau_0$,
\begin{eqnarray}
\bar{K}(x_2,p_2,t_2;x_1,p_1,t_1) = {\cal N}\int_{(x_1,p_1)}^{(x_2,p_2)}
\left[\frac{{\cal D}x{\cal D}p}{2\pi\hbar}\right] &\times&
e^{\frac{i}{\hbar}\int_{t_1}^{t_2}dt\left[\frac{1}{2}\left(\dot{x}p-x\dot{p}\right)
-\frac{1}{2m}p^2-\frac{1}{2}m\omega^2 x^2\right]}\times \nonumber \\
&\times& e^{-\frac{1}{\hbar}\int_{t_1}^{t_2}dt\frac{1}{2}\tau_0\left(m\omega\dot{x}^2+\frac{1}{m\omega}\dot{p}^2\right)},
\label{eq:Sc2}
\end{eqnarray}
which is well defined for $T_{21}>0$, while ${\cal N}$ is some normalisation factor to be specified shortly.
According to the discussion of Ref.\cite{Gov2.Klauder1}, we know that it ought to be possible to choose ${\cal N}$
in such a manner that
\begin{equation}
\lim_{\tau_0\rightarrow 0} \bar{K}(x_2,p_2,t_2;x_1,p_1,t_1)=K_0(x_2,p_2,t_2;x_1,p_1,t_1).
\label{eq:KK0}
\end{equation}

As a matter of fact, the kernel $\bar{K}(x_2,p_2,t_2;x_1,p_1,t_1)$ is the matrix element of the quantum
evolution operator defined by the action in (\ref{eq:Sc2}) in the two dimensional configuration space $(x,p)$
of that system, which is that of a particle moving in that Euclidean geometry and being subjected both to
a homogeneous magnetic field and some harmonic well. The Hilbert space of that system is certainly larger
than that of the original one dimensional harmonic oscillator, whose Hilbert space only coincides with the
lowest Landau level sector of the Landau problem defined by (\ref{eq:Sc2}). Hence one needs still to project
the above kernel $\bar{K}(x_2,p_2,t_2;x_1,p_1,t_1)$ onto its lowest Landau level sector, to reduce the quantum
dynamics for any finite value of $\tau_0$ to the Hilbert space of the original system, in order to keep to
a minimum any deformation of its quantum dynamics incurred by the nonvanishing time scale $\tau_0$.

In other words, when considering quantum states, $|\psi,t\rangle$, of the original quantum oscillator,
their time evolution is generated by a deformed quantum evolution operator $\hat{U}(t_2,t_1)$ such that
\begin{equation}
|\psi,t_2\rangle=\hat{U}(t_2,t_1)|\psi,t_1\rangle,
\end{equation}
where the operator $\hat{U}(t_2,t_1)$ is defined in terms of the original phase space coherent states by,
\begin{equation}
\hat{U}(t_2,t_1)=\int_{(\infty)}\frac{dx_2 dp_2}{2\pi\hbar}\int_{(\infty)}\frac{dx_1 dp_1}{2\pi\hbar}\,
|x_2,p_2\rangle\,\bar{K}(x_2,p_2,t_2;x_1,p_1,t_1)\,\langle x_1,p_1|.
\end{equation}
If the property (\ref{eq:KK0}) is indeed achieved, it is clear that one has,
\begin{equation}
\lim_{\tau_0\rightarrow 0}\hat{U}(t_2,t_1)=\hat{U}_0(t_2,t_1),
\end{equation}
meaning that in the limit $\tau_0\rightarrow 0$ the time evolution of the quantum Landau problem
projected onto its lowest Landau level reduces indeed to that of the original quantum system.

However, in contradistinction to the operator $\hat{U}_0(t_2,t_1)$, the deformed evolution operator
$\hat{U}(t_2,t_1)$ is neither reproducing nor unitary,
\begin{equation}
\hat{U}(t_3,t_2)\hat{U}(t_2,t_1)\ne \hat{U}(t_3,t_1),\qquad
\hat{U}^\dagger(t_2,t_1)\ne \hat{U}(t_1,t_2).
\end{equation}
The lack of unitarity follows directly from the real exponential Gaussian factor involving $\tau_0$,
namely the Brownian motion and thus statistically stochastic character associated to that term, while the same
term implies some form of decoherence in the dynamics of the system which entails a lack
of reproducibility under convoluted time evolution. The deformation parameter $\tau_0$ is thus directly responsible for the
existence of an arrow of time at the fundamental quantum level. However, any deviations from the ordinary unitary and
reversible quantum dynamics is expected to remain extremely small provided the fundamental constant
$\tau_0$ remains much smaller than any other time scale of the undeformed quantum dynamics. A time scale
on the order of the Planck time is a clear candidate.

\subsection{The deformed time dynamics}

All that remains to be done is the explicit evaluation of the path integral (\ref{eq:Sc2}) of
the extended Landau problem. This may be done through different and complementary approaches.
Since the extended action remains quadratic in all $(x,p)$ variables, a saddle-point evaluation
is a clear possibility, which must lead to an expression of the form
\begin{equation}
\bar{K}(x_2,p_2,t_2;x_1,p_1,t_1)=\bar{K}(T_{21})\,e^{-\frac{1}{\hbar}S^{\rm extended}_c},
\label{eq:Kpre}
\end{equation}
where $S^{\rm extended}_c$ is the value of the extended action
\begin{equation}
S^{\rm extended}[x,p]=\int_{t_1}^{t_2}dt\left[\frac{1}{2}\tau_0\left(m\omega\dot{x}^2+\frac{1}{m\omega}\dot{p}^2\right)
-\frac{1}{2}i\left(\dot{x}p-x\dot{p}\right)-\frac{i}{2m}p^2-\frac{1}{2}im\omega^2x^2\right],
\label{eq:Sext}
\end{equation}
for the classical trajectory associated to the boundary values $(x(t_1),p(t_1))=(x_1,p_1)$ and
$(x(t_2),p(t_2))=(x_2,p_2)$. Even though these boundary values are real, since the equations of
motion following from (\ref{eq:Sext}) are complex one needs to consider the analytic continuation
of the dynamics defined by this extended action to the complexification of its four dimensional phase
space. Nevertheless, such an evaluation of the path integral (\ref{eq:Sc2}) through complex analytic
continuation is in perfect accordance with the usual analytic continuation methods for the integration of
ordinary functions. The evaluation of the exponential factor in (\ref{eq:Kpre}) is thus simply a matter
of solving second order linear differential equations.

For what concerns the prefactor $\bar{K}(T_{21})$ in (\ref{eq:Kpre}), its final evaluation may be
determined by imposing the reproducing property that the kernel $\bar{K}(x_2,p_2,t_2;x_1,p_1,t_1)$
must obey as the evolution operator of the extended Landau problem in its entire Hilbert space of states,
including all its possible Landau levels. This requirement determines the prefactor only up to an
arbitrary factor of the form
\begin{equation}
e^{\gamma(\tau_0) T_{21}},
\end{equation}
where $\gamma(\tau_0)$ is some unspecified function of $\tau_0$. As a matter of fact, the imaginary
contribution to this function is in direct relation with the quantum vacuum energy of the original
quantum oscillator, and may thus be chosen accordingly in order to meet the property (\ref{eq:KK0})
as is also required in the Daubechies--Klauder construction\cite{Gov2.Klauder1}. For what concerns the
real part of $\gamma(\tau_0)$, this must be chosen such that the $\tau_0\rightarrow 0$ limit of
$\bar{K}(x_2,p_2,t_2;x_1,p_1,t_1)$ remains finite and nonvanishing. Finally, note that these
considerations should also remain consistent with the fact that
\begin{equation}
\lim_{T_{21}\rightarrow 0}\bar{K}(x_2,p_2,t_2;x_1,p_1,t_1)=2\pi\hbar\,\delta(x_2-x_1)\delta(p_2-p_1),
\end{equation}
since $\bar{K}(x_2,p_2,t_2;x_1,p-1,t_1)$ is the evolution kernel of the extended Landau problem in its two dimensional
configuration space $(x,p)$.

Given the explicit evaluation of (\ref{eq:Kpre}) along such lines\cite{Gov2.Matt}, there remains only to substitute
that result in the definition of the deformed quantum evolution operator $\hat{U}(t_2,t_1)$ for the
original oscillator system, whose dynamics is confined to the lowest Landau level of the extended dynamics.
In particular, its phase space coherent state matrix elements are given by the kernel,
\begin{equation}
K(x_2,p_2,t_2;x_1,p_1,t_1)=\langle x_2,p_2|\hat{U}(t_2,t_1)|x_1,p_1\rangle .
\end{equation}
As remarked previously, for any finite value of $\tau_0$ this deformed kernel is neither reproducing
nor unitary, while in the limit $\tau_0\rightarrow 0$ it reduces to the kernel $K_0(x_2,p_2,t_2;x_1,p_1,t_1)$
of the undeformed oscillator in its Hilbert space.

In order to provide a closed form expression for the deformed kernel $K(x_2,p_2,t_2;x_1,p_1,t_1)$,
let us introduce the following notations. First we have
\begin{equation}
R^2_0=\sqrt{1+16\omega^2\tau^2_0},\quad R_0>0,\qquad
R=\sqrt{\frac{1}{2}(R^2_0+1)},\qquad
S=\frac{1}{2}(R+1),
\end{equation}
which are such that
\begin{equation}
R-1=2\frac{\omega^2\tau^2_0}{R^2S},\qquad
\frac{\omega\tau_0}{RS}=\sqrt{1-S^{-1}}.
\end{equation}
Note that in the limit $\tau_0\rightarrow 0$ all these quantities, $R_0$, $R$ and $S$,
reduce to a unit value. Next let us define
\begin{eqnarray}
F(T_{21}) &=& \frac{1}{S}\frac{R^2+2i\omega\tau_0}{R+2i\omega\tau_0}
\frac{1}{1+\frac{R-1}{R+1}\frac{R-2i\omega\tau_0}{R+2i\omega\tau_0}
e^{-\frac{R}{\tau_0}T_{21}} e^{-2i\frac{\omega}{R}T_{21}}} \nonumber \\
&=& \left[e^{-\frac{R}{\tau_0}T_{21}}e^{-2i\frac{\omega}{R}T_{21}}
+S\frac{R+2i\omega\tau_0}{R^2+2i\omega\tau_0}
\left(1-e^{-\frac{R}{\tau_0}T_{21}}e^{-2i\frac{\omega}{R}T_{21}}\right)\right]^{-1} .
\end{eqnarray}
In the limit $\tau_0\rightarrow 0$ this quantity also reduces to a unit value.

The explicit evaluation of the deformed kernel $K(x_2,p_2,t_2;x_1,p_1,t_1)$ then finds,
\begin{equation}
K(x_2,p_2,t_2;x_1,p_1,t_1) = \ \ e^{-\frac{1}{2}|z_2|^2-\frac{1}{2}|z_1|^2}\,
e^{-\frac{1}{2}i\frac{\omega}{R}T_{21}}\
F(T_{21})\,\exp\left[F(T_{21})\,e^{-(R-1)\frac{T_{21}}{2\tau_0}}\,
e^{-i\frac{\omega}{R}T_{21}}\,\bar{z}_2 z_1\right],
\end{equation}
with $z_i=\sqrt{m\omega/(2\hbar)}(x_i+ip_i/(m\omega))$ ($i=1,2$).
In this form it is quite clear that in the limit $\tau_0\rightarrow 0$ this expression reduces
indeed to that for the undeformed kernel, $K_0(x_2,p_2,t_2;x_1,p_1,t_1)$.

Before considering the physical properties implied by this result, let us also give the
time evolution of Fock states implied by the deformed quantum dynamics,
\begin{equation}
K_{n,\ell}(t_2,t_1)\equiv\langle n,t_2|\ell,t_1\rangle\equiv \langle n|\hat{U}(t_2,t_1)|\ell\rangle=
\delta_{n,\ell}\,F^{n+1}(T_{21})\,e^{-n\frac{R-1}{2\tau_0}T_{21}}\,
e^{-i\frac{\omega}{R}\left(n+\frac{1}{2}\right)T_{21}}.
\end{equation}
Note that this Fock space kernel is diagonal in the Fock basis, which is a welcome result since the deformation
thus does not lead to a mixing of Fock states. Furthermore, this kernel is also such that, as it should,
\begin{equation}
\lim_{\tau_0\rightarrow 0}K_{n,\ell}(t_2,t_1)=\delta_{n,\ell}\,
e^{-i\omega\left(n+\frac{1}{2}\right) T_{21}},\qquad
\lim_{T_{21}\rightarrow 0^+}K_{n,\ell}(t_2,t_1)=\delta_{n,\ell},
\end{equation}
while except for the factor $F^{n+1}(T_{21})$, that kernel displays a p\^ole
structure in the complex ``energy" or frequency plane given by the expression
\begin{equation}
\pm\left[\frac{\omega}{R}-i\frac{R-1}{2\tau_0}\right]=\pm\frac{\omega}{R}
\left[1-i\sqrt{1-S^{-1}}\right].
\end{equation}
For a finite value of $\tau_0$ the values of $\pm\omega$ are thus displaced
inside the complex plane in precisely the same manner as implied by the $+i\epsilon$ prescription
used in quantum field theory in order to construct the time causal Feynman propagator.
In the present context however, the shift by an imaginary quantity remains finite, with
as consequence a decohering, non unitary and yet quantum time evolution.

Finally let us consider the time evolution of an arbitrary state, $|\psi,t\rangle$.
From its decomposition in the Fock basis and given an initial configuration, $|\psi,t_1\rangle$, one finds
\begin{equation}
|\psi,t_2\rangle=\sum_{n=0}^\infty\,|n\rangle\,\psi_n(t_2),
\end{equation}
where the coefficients $\psi_n(t_2)$ are such that the occupation numbers for each of
the Fock states evolve according to
\begin{equation}
\frac{|\psi_n(t_2)|^2}{|\psi_n(t_1)|^2}=
P^{n+1}_0\,\left(1+2\frac{R-1}{R+1}\cos 2\left(\frac{\omega}{R}T_{21}+\varphi\right)e^{-\frac{R}{\tau_0}T_{21}}
+\left(\frac{R-1}{R+1}\right)^2 e^{-2\frac{R}{\tau_0}T_{21}}\right)^{-(n+1)}\
e^{-n\frac{R-1}{\tau_0}T_{21}},
\end{equation}
where
\begin{equation}
\tan\varphi=2\frac{\omega\tau_0}{R},\qquad
P_0=\left|\frac{1}{S}\frac{R^2+2i\omega\tau_0}{R+2i\omega\tau_0}\right|^2=
\frac{1}{S^2}\frac{R^4+4\omega^2\tau^2_0}{R^2+4\omega^2\tau^2_0}.
\end{equation}
As it should, in the absence of $\tau_0$ each of these occupation numbers are conserved quantities,
whereas in presence of $\tau_0$ they have a complicated time dependence, even though they all
decouple exponentially in succession starting with the higher Fock states as time grows larger,
leaving over only the Fock vacuum at $n=0$ with a finite occupation determined by the value of $P_0$.
More precisely, for each value of $n\ne 0$, the overall exponentially decreasing behaviour governed by the time scale
$\tau_0/(n(R-1))$ is modulated by the prefactor $|F(T_{21})|^{2(n+1)}$ which besides its own exponentially
decreasing behaviour towards the value $P^{n+1}_0$ governed by the time scale $\tau_0/R$ also
oscillates with a periodicity $\pi R/\omega$.

All these expressions display the fact that for any finite value of the fundamental time scale $\tau_0$,
in combination with the proper time scale $2\pi/\omega$ of the oscillator the time evolution of the deformed
quantum oscillator is characterised by an oscillatory time scale $\pi R/\omega$ and
two further exponentially decreasing time scales $\tau_-$ and $\tau_+$ such that,
\begin{equation}
\tau_-=2\frac{\tau_0}{R} < \tau_+=2\frac{\tau_0}{R-1}=\frac{R^2 S}{\omega^2\tau_0},
\end{equation}
the smaller of the two vanishing and the larger one diverging linearly with $\tau_0$ as $\tau_0\rightarrow 0$.
Furthermore, as a function of the ratio of the oscillator intrinsic time scale, $1/\omega$,
with the fundamental time scale, $\tau_0$, namely the quantity $1/(\omega\tau_0)$, one has
(see Figs.~1 and 2),
\begin{equation}
\frac{1}{\omega\tau_0}\rightarrow 0\ :\qquad
\tau_-\simeq\tau_0\sqrt{\frac{2}{\omega\tau_0}},\qquad
\tau_+\simeq\tau_0\sqrt{\frac{2}{\omega\tau_0}},\qquad
P_0\simeq\frac{4}{\omega\tau_0},
\end{equation}
while
\begin{equation}
\frac{1}{\omega\tau_0}\rightarrow\infty\ :\qquad
\tau_-\simeq 2\tau_0,\qquad
\tau_+\simeq \frac{1}{\omega^2\tau_0},\qquad
P_0\simeq 1.
\end{equation}
In actual fact, $P_0$ reaches a single maximum value exceeding unity, $P_0\simeq 1.079$,
for $1/(\omega\tau_0)\simeq 2.591$ (for further graphs of different quantities, see Ref.\cite{Gov2.Matt}).

\begin{figure}[h!]
\begin{center}
\includegraphics[angle=0, width=6 cm,height=4.5 cm  ]{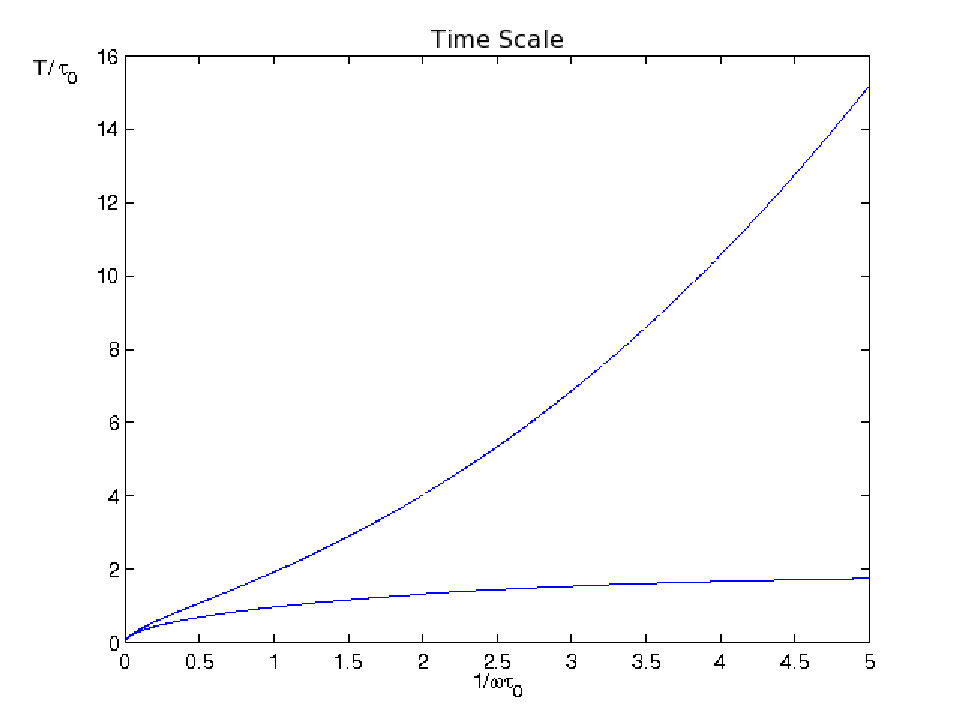}
\caption[]{The ratios $\tau_+/\tau_0$ (upper curve) and $\tau_-/\tau_0$
(lower curve) as a function of $1/(\omega\tau_0)$.}
\end{center}
\end{figure}

\begin{figure}[h!]
\begin{center}
\includegraphics[angle=0, width=6 cm,height=4.5 cm ]{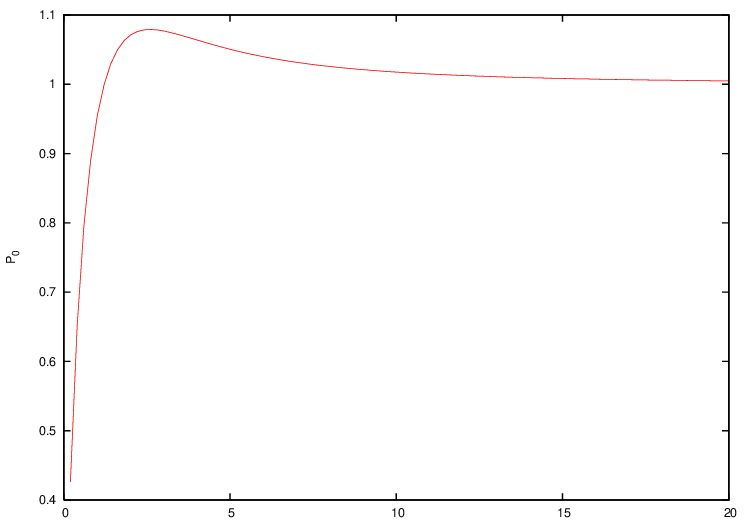}
\caption[]{The Fock vacuum asymptotic occupation number $P_0$ as a function
of $1/(\omega\tau_0)$.}
\end{center}
\end{figure}

The regime of ordinary quantum mechanical behaviour is thus reached in the limit $1/(\omega\tau_0)\rightarrow \infty$,
namely when the proper time scale of the system is much larger than that of the fundamental scale $\tau_0$.
More precisely, even if these two scales differ by two to three orders of magnitude only,
namely $\frac{1}{\omega\tau_0}\simeq 10^2-10^3$, deviations from
ordinary quantum behaviour are already small since $R_0$ differs from unity by a term of order $(\omega\tau_0)^2$.
When moving closer to time scales $1/\omega$ becoming comparable to $\tau_0$, three main regions in time evolution need to
be distinguished (see Figs. 1 and 2). When $0<T_{21}<\tau_-$, the behaviour of the occupation numbers is dominated by the time dependence
of $|F(T_{21})|^2$ which besides an amplitude exponentially reaching the value of $P_0$, also carries an
oscillatory pattern set by the angular frequency $2\omega/R<2\omega$. Once the region $\tau_-<T_{21}<\tau_+$ is
reached, the exponential decrease of time scale $\tau_+$ dominates the time dependence, which converges towards
a totally collapsed and decoherent behaviour leaving only the Fock vacuum at $n=0$ occupied in the region $T_{21}>\tau_+$.
Given a fixed intrinsic time scale $1/\omega$ of the system, the smaller the constant $\tau_0$,
the wider the interval $[\tau_-,\tau_+]$ with $\tau_-$ moving closer to a vanishing value, and the lesser
the observational effects of the deformation parameter $\tau_0$ over an appreciable length of time.
However the larger $\tau_0$, the tighter becomes the interval $[\tau_-,\tau_+]$ and the sooner one enters the collapsed
decoherent regime leaving the system in its ground state with no dynamics. The consequences of the
interfering and competing time dependent effects of both quantum phases and Brownian motion diffusion in phase space are thus quite
subtle and rich, leading to a tantalising picture for such a deformed quantum dynamics.

In the absence of any intrinsic time scale for the system, no observable effects related to $\tau_0$ are possible.
For instance, taking the limit of a vanishing angular frequency $\omega$, corresponding to the one dimensional
free nonrelativistic massive particle, and considering the configuration space evolution kernel
$K(x_2,t_2;x_1,t_1)=\langle x_2|\hat{U}(t_2,t_1)|x_1\rangle$, one finds
\begin{equation}
\lim_{\omega\rightarrow 0}K(x_2,t_2;x_1,t_1)=
\left(\frac{m}{2i\pi\hbar T_{21}}\right)^{1/2}\,e^{i\frac{m}{2\hbar T_{21}}(x_2-x_1)^2},
\end{equation}
which coincides with the ordinary evolution kernel for that system, irrespective of the value for $\tau_0$.

Finally, coming back to the issue of noncommutativity, in the present system only the commutator of the
$\hat{x}$ and $\hat{p}$ operators in the deformed Heisenberg picture may be considered. Using for the definition of quantum
operators in the Heisenberg picture the time evolution generated by the deformed operator $\hat{U}(t_2,t_1)$,
one finds
\begin{eqnarray}
\left[\hat{x}(t_2),\hat{p}(t_2)\right] &=& i\hbar\,
\Big\{ |\Omega_0\rangle\,|F(T_{21})|^6 e^{-\frac{R-1}{\tau_0}T_{21}}\langle\Omega_0|\,+\,\nonumber \\
&&+\sum_{n=1}^\infty|n\rangle\left[(n+1)|F(T_{21})|^4 e^{-2\frac{R-1}{\tau_0}T_{21}}-n\right]
|F(T_{21})|^{2(2n+1)} e^{-(2n-1)\frac{R-1}{\tau_0}T_{21}}\langle n|\Big\}.
\end{eqnarray}
In the limit $\tau_0\rightarrow 0$, one indeed recovers the Heisenberg algebra. However for a finite value of $\tau_0$,
the algebra is indeed deformed, but in a far more complicated and in a time dependent manner than displayed by
the forms of noncommutativity mentioned in the Introduction. In particular, note that in the limit $T_{21}\rightarrow\infty$,
the above commutator vanishes identically, which is consequence of the fact that the system then collapses to its ground state
as the only state being occupied, its Hilbert thus becoming effectively one dimensional.

Incidentally, when extended to a two dimensional harmonic oscillator, one may also consider in the deformed Heisenberg
picture the commutator of the two cartesian coordinate operators, $\hat{x}_1(t_2)$ and $\hat{x}_2(t_2)$, to find\cite{Gov2.Matt}
that in contradistinction to the ordinary case, this commutator is no longer vanishing, even though its expression
is quite more involved than simply being a nonvanishing constant as applies to the cases mentioned in the
Introduction.

\section{Prospects and Speculations}
\label{Gov2.Sec5}

By taking seriously the suggestion that in the Daubechies--Klauder construction of the phase space coherent state path integral
the regularisation parameter $\tau_0$ may in fact be a new fundamental physical constant,
some tantalising features have come to the fore, even if only for as simple a system as the one dimensional
harmonic oscillator. Extrapolating these results to the quantum field theory context, albeit in a perturbative
regime for interactions, one is led to the conclusion that at the very least this construction provides a
definition of deformed quantum field theories void of any short distance singularities. Indeed, for modes of
arbitrarily large frequency in comparison to $\tau_0$, in no time whatever do these mode reach their totally
decoherent and collapsed ground state. Short distance singularities are smoothed out simply because for those
distance scales still smaller than that defined by $\tau_0$, $\hbar$ and $c$, no field dynamics is available
to probe the geometry of spacetime at such scales. To an observer, the geometry of spacetime would appear coarse grained,
with an effective exponential cut-off set essentially by the value of $\tau_0$. Even a quantum field theory of General Relativity
would remain finite at short distances, with its classical singularities smoothed out within the present formalism of a
deformed quantum physics.

In a gravitational and cosmological context, such a proposal raises some intriguing possibilities.
Not only would the initial Big Bang singularity be smoothed out, but the fact that the occupation
numbers of different field modes would evolve in time at different rates would also bear consequences
for the interpretation of the microwave background and for the observed accelerated expansion 
of the Universe, the latter having led to the hypothesis of dark energy. If indeed light gets ``tired" in the manner
described above but also modulated by a function of its frequency and thus of its red-shift, such conclusions would need to be reassessed.
In the same spirit, in the context of black hole physics, back-reaction of the transplanckian modes on
Hawking's radiation could be addressed anew, paying due attention also to the decohering effects at 
the horizon. And there is of course also the question of the eventual relation between the time scale $\tau_0$ and Planck's time scale,
or equivalently, Newton's gravitational constant $G_N$.

On a more practical level, not only based on the above considerations but possibly also through
precision studies of atomic or nuclear spectra, especially for metastable states, it may be possible to
identify stringent upper bounds on the value for the constant $\tau_0$. Another issue is how a finite
$\tau_0$ affects the classical limit $\hbar\rightarrow 0$.

Finally, within a relativistic quantum field theory context, it is clear that a finite $\tau_0$ is a source
of violation of Lorentz symmetry, besides a fundamental breaking of time reversal symmetry and lack of quantum unitarity.
However over recent years one has grown accustomed with the possibility that within a theory for quantum gravity and
quantised spacetime geometry, it may well be the case that such features must be accepted on their own merits,
and that for instance Lorentz symmetry may no longer need to be a fundamental invariance of spacetime.

Hence much remains to be explored to assess the possible physics merits of the proposal made in the
present contribution towards a specific deformation of quantum dynamics, motivated by a particular
regularised construction of the phase space path integral due to Ingrid Daubechies and John Klauder,
and within the perspective of noncommutative geometries of spacetime for a quantum theory of gravity
unified with all other three fundamental interactions.

\section*{Acknowledgements}

J.G. wishes to address here warm words of appreciation to Prof. John Klauder for constructive discussions,
his interest in the present work, and his support for many years.
J.G. acknowledges the Abdus Salam International Centre for Theoretical Physics (ICTP, Trieste, Italy)
Visiting Scholar Programme in support of a Visiting Professorship at the ICMPA-UNESCO (Republic of Benin).
This work is supported by the Institut Interuniversitaire des Sciences Nucl\'eaires, and by
the Belgian Federal Office for Scientific, Technical and Cultural Affairs through
the Interuniversity Attraction Poles (IAP) P6/11.

\end{document}